\documentclass[aps,prd,preprintnumbers,superscriptaddress,nofootinbib,floatfix,twocolumn,notitlepage]{revtex4-2}
\usepackage{graphicx}
\usepackage{dcolumn}
\usepackage{bm}
\usepackage[normalem]{ulem} 
\usepackage{amsmath,amssymb,mathtools}
\usepackage{slashed}
\usepackage{mathrsfs}
\usepackage{array}
\usepackage{textcomp}
\usepackage{xcolor}
\usepackage[colorlinks=true,citecolor=blue,urlcolor=blue,linktocpage=true,linkcolor=blue]{hyperref}
\usepackage{siunitx}
\usepackage{booktabs}
\usepackage{tabularx}
\usepackage{multirow}
\usepackage{xspace}
\usepackage{physics}
\usepackage{dsfont}
\usepackage{physunits}
\usepackage{ragged2e}
\usepackage{url}     
\usepackage{cancel}
\usepackage{subcaption}  
\newcommand{\bea}{\begin{eqnarray}}
\newcommand{\eea}{\end{eqnarray}}

\newcommand{\Bsm}{\baryon}

\newcommand{\ie }{\textit{i.e.}}
\newcommand{\cY}{\mathcal{Y}}
\newcommand{\cL}{\mathcal{L}}

\newcommand{\cO}{\mathcal{O}}
\newcommand{\cM}{\mathcal{M}}

\newcommand{\GeV}{\textrm{GeV}}
\newcommand{\TeV}{\textrm{TeV}}
\newcommand{\baryon}{\mathfrak{B}}

\newcommand{\YBAU}{Y_{\rm obs}}
\newcommand{\Ybar}{Y_{\baryon}}
\newcommand{\beq}{\begin{equation}}
\newcommand{\eeq}{\end{equation}}
\def\beqa{\begin{eqnarray}}
\def\eeqa{\end{eqnarray}} 
\newcommand{\Br}[1]{\text{BR}\left(#1\right)}
\newcommand{\BR}{\textrm{BR}}
\newcommand{\MET}{\slashed{E}}
\newcommand{\ACP}{ A^{\cM}_{i}}
\newcommand{\ACPdir}{ A^{\cM}_{\baryon,{\rm dir}}}

\usepackage{tikz-feynman}
\usepackage[dvipsnames]{xcolor}

\begin{document}

\preprint{CERN-TH-2026-071}
\title{$\baryon$locking Mesogenesis}

\author{Chaja Baruch}
\email{chajabaruch@campus.technion.ac.il}
\affiliation{Physics Department, Technion--Israel Institute of Technology, Haifa 3200003, Israel}
\author{Gilly Elor}
\email{gilly.elor@austin.utexas.edu}
\affiliation{Theory Group, The Weinberg Institute for Theoretical Physics, University of Texas at Austin, Austin, TX 78712, United States}
\author{Jared M. Goldberg}
\email{jmgoldberg@campus.technion.ac.il}
\affiliation{Physics Department, Technion--Israel Institute of Technology, Haifa 3200003, Israel}
\author{Omer Shtaif}
\email{omer.shtaif@campus.technion.ac.il}
\affiliation{Physics Department, Technion--Israel Institute of Technology, Haifa 3200003, Israel}
\author{Yotam Soreq}
\email{soreqy@physics.technion.ac.il}
\affiliation{Physics Department, Technion--Israel Institute of Technology, Haifa 3200003, Israel}
\affiliation{Theoretical Physics Department, CERN, 1211 Geneva 23, Switzerland}

\begin{abstract}
Mechanisms of Mesogenesis generate the baryon asymmetry and dark matter of the Universe through late-time decays of Standard Model mesons into baryons and dark matter states. 
Utilizing the CP violation in the meson systems themselves, the resulting baryon asymmetry is directly controlled by collider observables, the CP asymmetry $A_{CP}$, and the branching fraction for the meson decays. 
Experimental probes of these decays place strong constraints on the amount of $CP$ violation required, placing it well above observed limits in meson mixing. 
Additionally, strong lower bounds on the proton lifetime seemingly rule out Mesogenesis mechanisms which use $D$ mesons. 
In this work, we propose to circumvent these constraints by ``morphing'' the mass of the dark sector particles using a late-time phase transition. 
The change in mass of the final decay products kinematically excludes meson and proton decays, relaxing the constraints on the model parameter space. 
\end{abstract}

\maketitle

\section{Introduction }
\label{sec:intro}

The origin of the observed baryon asymmetry of the Universe~(BAU) remains one of the most outstanding mysteries in particle physics.
Measurements of the Cosmic Microwave Background~(CMB)~\cite{Ade:2015xua,Aghanim:2018eyx} and light element abundances after Big Bang nucleosynthesis~(BBN)~\cite{Cyburt:2015mya,ParticleDataGroup:2024cfk}, determine the BAU to be:   
\begin{align}
    \label{eq:YBAU}
    \YBAU 
    \equiv 
    (n_\baryon-n_{\bar{\baryon}})/s 
    = 
    \left( 8.72 \pm 0.04 \right) \times 10^{-11}\,,
\end{align}
where $n_\baryon\,(n_{\bar{\baryon}})$ is the baryon\,(anti-baryon) number density and $s$ is the Universe entropy density.
Different mechanisms have been proposed to explain the origin of the BAU; for example, Electroweak Baryogenesis~\cite{Kuzmin:1985mm,Shaposhnikov:1987tw,Cohen:1990py} and Leptogenesis~\cite{Fukugita:1986hr,Davidson:2008bu} (see~\cite{Falkowski:2017uya} for relation with light dark matter).  
See Ref.~\cite{Elor:2022hpa} for reviews. 
Typically, these mechanisms predict high-scale new physics, which is challenging to test directly in current experiments. 

\begin{figure*}[t]
    \centering
\resizebox{\textwidth}{!}{%
\begin{tikzpicture}[
  x=2.2cm,y=2.31cm,
  >=Latex,
  line width=0.8pt,
  every node/.style={transform shape, scale=1.1},
  event/.style={circle,draw,fill=white,inner sep=1.0pt},
  eventred/.style={circle,draw=red!70!black,fill=red!70!black,inner sep=1.0pt},
  eventblue/.style={circle,draw=blue!70!black,fill=blue!70!black,inner sep=1.0pt},
  lab/.style={font=\scriptsize,align=center,text width=2.45cm},
  ticklab/.style={font=\scriptsize},
  lab/.style={font=\scriptsize,align=center,text width=2.45cm},
  ticklab/.style={font=\scriptsize},
]

\def\yAxis{0}
\def\yTop{0.7}
\def\yBot{-0.7}

\def\xmin{0}
\def\xmax{7.0}

\def\xQCD{0.9}
\def\xMeso{2.925}
\def\xPT{5.125}
\def\xBBN{5.9}

\def\xmesoleft{1.2}
\def\xmesoright{4.65}
\def\blockingleft{4.65}
\def\blockingright{5.6}
\draw[
  blue!70!black,
  <->,
  double=blue!70!black,
  double distance=0.6pt,
  line width=0.2pt
] (\blockingleft,-0.08) -- (\blockingright,-0.08);
\draw[
  red!70!black,
  <->,
  double=red!70!black,
  double distance=0.6pt,
  line width=0.2pt
] (\xmesoleft,0.08) -- (\xmesoright,0.08);
\draw[->] (\xmin,\yAxis) -- (\xmax,\yAxis);
\node[ticklab,anchor=south west] at (\xmin,\yAxis) {hot / early};
\node[ticklab,anchor=south east] at (\xmax,\yAxis) {cooler / later};

\foreach \x/\t in {
  \xQCD/$\sim150$ MeV,
  \xBBN/$\sim 5$ MeV
}{
  \draw (\x,\yAxis) -- (\x,\yAxis-0.09);
  \node[ticklab,anchor=north] at (\x,\yAxis-0.12) {\t};
}

\node[event] (qcd) at (\xQCD,\yAxis) {};
\draw (qcd) -- (\xQCD,\yTop-0.1);
\node[lab,anchor=center] at ($(\xQCD,\yTop)+(0,0.06)$)
{\textbf{QCD Phase Transition}};

\node[eventred] (meso) at (\xMeso,\yAxis+0.06) {};
\draw[red!70!black,dashed] (meso) -- (\xMeso,\yTop-0.1);
\node[
  lab,
  anchor=center,
  fill=red!12,
  draw=red!60!black,
  rounded corners=1pt,
  inner sep=2pt,
  text width=5.cm      
] at ($(\xMeso,\yTop)-(0,0)$)
{\textbf{Mesogenesis window}\\
Generation of BAU\\[6pt]
$\begin{array}{c|c}
   B\to\baryon\psi  &  D\to\baryon\psi \\[6pt]
   1<m_\psi\lesssim 2.5\GeV & 0<m_\psi\lesssim 0.6\GeV
\end{array}$};

\node[eventblue] (pt) at (\xPT,\yAxis-0.06) {};
\draw[blue!70!black,dashed] (pt) -- (\xPT,\yBot+0.08);
\node[
  lab,
  anchor=center,
  fill=blue!10,
  draw=blue!60!black,
  rounded corners=1pt,
  inner sep=2pt,
  text width=7cm
    ] at ($(\xPT,\yBot)+(0,0)$)
{\textbf{Blocking window}\\[6pt]
$\begin{array}{c | c}

B\to\baryon\psi
&
D\to\baryon\psi \\[6pt]

\Delta m_\psi > \max\!\left[m_{\mathcal M}-m_\baryon-m_\psi\right]\,\,
&
\,\,\Delta m_\psi > m_p - m_\psi
\end{array}$
};

\node[event] (bbn) at (\xBBN,\yAxis) {};
\draw (bbn) -- (\xBBN,\yTop-0.1);
\node[lab,anchor=center] at ($(\xBBN,\yTop)+(0.0,0)$)
{\textbf{BBN}};

\end{tikzpicture}
}
\caption{The thermal history of the Universe from QCD to BBN -- the window for Mesogenesis. }
\label{fig:timeline}
\end{figure*}
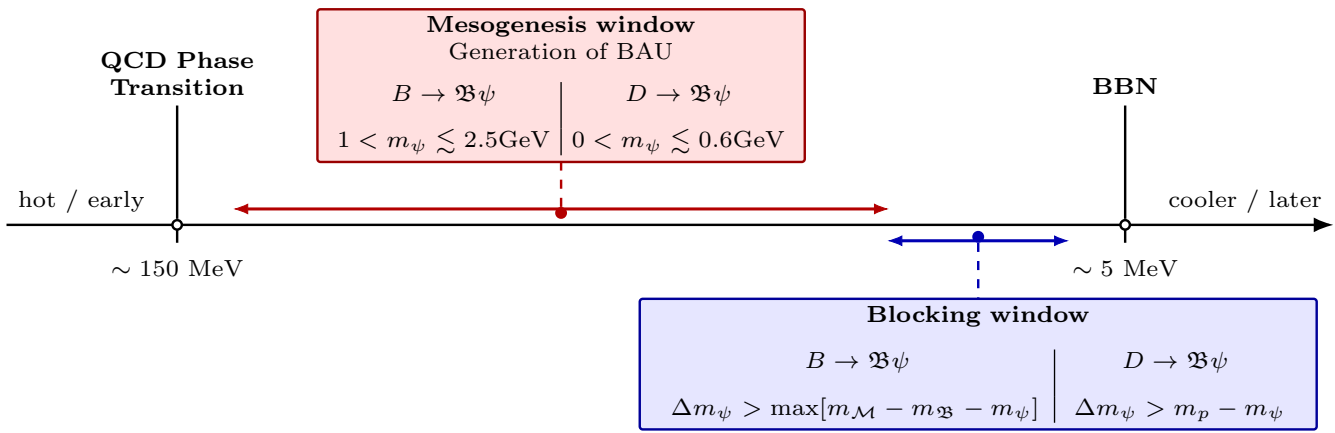

In contrast, \textit{Mesogenesis}~\cite{Aitken:2017wie,Elor:2018twp,Elor:2025fcp, Elor:2024cea, Elor:2020tkc,Elahi:2021jia} has been proposed as a testable paradigm of late-time and low-scale baryogenesis, where the BAU is generated through Standard~Model~(SM) meson ($\cM$) decays into a SM~baryon ($\baryon$) and dark-matter states via an unstable dark baryon-number carrying fermion ($\psi$).
Mesogenesis models necessarily assume a radiation-dominated universe in which the matter density is dominated by a scalar field, $\Phi$, of mass $m_\Phi > 2 m_{\cM}$.
$\Phi$ decays to quarks and anti-quarks after the QCD phase transition, $T\sim150\,\MeV$, and before BBN, $T\sim5\,\MeV$.
See also Fig.~\ref{fig:timeline}.

Within the Mesogenesis framework, the ratio of predicted to observed BAU can be written as~\cite{Elor:2025fcp, Elahi:2021jia, Elor:2018twp,Alonso-Alvarez:2021qfd,Alonso-Alvarez:2019fym}
\begin{align}
    \label{eq:YBmeso}
    \frac{\Ybar}{\YBAU}
    \simeq 
    \sum_{\cM,\baryon,i} 
    \frac{
    \BR(\cM \to \baryon  \psi ) \times \, 
    \ACP \times \alpha_{i}(T_R)}{ 1.7\times 10^{-6}} \, ,  
\end{align}
where the indices $\cM$, $\baryon$ and $i$ run over the possible mesons, baryons and CP asymmetries, respectively. 
$\BR(\cM\to\baryon\psi)$ is the relevant branching ratio; $\ACP$ is the CP asymmetry in the specific mode, which can be sourced from either direct CP in the decay or, for neutral mesons also in the mixing.
The function $\alpha_{i}(T_R)$ encodes the cosmological evolution of the baryon-antibaryon density difference, evaluated at the reheating temperature, $T_R$, defined by the $\Phi$ decay rate as $\Gamma_\Phi \equiv 3H(T_R)$. As reviewed in prior Mesogenesis constructions ~\cite{Aitken:2017wie,Elor:2018twp,Elor:2025fcp, Elor:2024cea, Elor:2020tkc,Elahi:2021jia}, we will employ models in which $\psi$ promptly decays into a pair of  stable dark-sector particles, discussed in the following section, in order to prevent washout from $\psi$ decays back to the SM.

Mesogenesis models face two important challenges. 
First, even if the SM-dark sector mediator is not coupled directly to light quarks, proton decay is induced at least at loop level unless $\psi$ is heavier than the proton, i.e. $m_\psi > m_p$.
However, for $D$ Mesons, the phase space for the decay $D\to \baryon \psi$ requires $m_\psi \leq m_D-m_{\Bsm} \lesssim 0.92\,\GeV$, where the upper bound is obtained for the lightest possible baryon.
Thus, the only existing models of $D$-Mesogenesis necessarily involve light dark sector lepton states~\cite{Elor:2020tkc}, and not baryons. 

Second, from Eq.~\eqref{eq:YBmeso}, reproducing the observed BAU requires $\BR(\cM\to\baryon\psi)\, A_i^\cM\, \alpha_i\sim 10^{-6}$.
For CP violation~(CPV) from meson mixing, $\alpha_{i}\lesssim1$ and $|A_{{\rm mix}}^{B_s}|\lesssim 2\times 10^{-3}$~\cite{HeavyFlavorAveragingGroupHFLAV:2024ctg}.
When the observed upper bound on CPV in mixing is saturated, this constrains the branching fraction $\BR(\cM\to \baryon\psi)\gtrsim 10^{-3}$. 
A stronger constraint is obtained by considering the SM prediction for the CPV in meson mixing, in which case the required branching fraction is increased to $\sim 0.1$. 
However, dedicated searches for $\cM\to \baryon \MET$ place stringent bounds of $\BR(B\to\baryon\psi)\lesssim 10^{-5}$ for specific modes such as $B^0\to\Lambda\MET$~\cite{Belle:2021gmc,BaBar:2023rer}, $B^+\to p\MET$~\cite{BaBar:2023dtq} and $B^+\to \Lambda^+_c\MET$~\cite{BaBar:2024qqx}.
One concludes that this mechanism is under tension, in particular if only SM-CPV in $B^0$ and $B^0_s$ is considered.

For direct CPV in Meson decays, $\alpha_i\sim\cO(100)$~\cite{Elor:2025fcp}, meaning the BAU can be explained by $\BR(\cM\to\baryon\psi)\, A_{\rm dir}^{\cM}\sim10^{-8}$.
While there are no bounds on $A^\cM_{\rm dir}$, it is expected to be suppressed by a loop factor or small strong phases, relaxing the requirement on the branching fraction to $\BR(\cM\to\baryon\psi)\sim10^{-5}$.
Thus, Mesogenesis from direct CPV can be under tension for decay modes with sensitivity of $10^{-5}-10^{-6}$, which is within current experimental reach.
For probes other than $\cM \to \baryon\MET$, such as collider constraints, see Ref.~\cite{Alonso-Alvarez:2021qfd,Hiller:2026osz}. 

The above tensions are based on the assumption that $\cM\to\baryon\psi$ (or the proton lifetime) today is unchanged from its early Universe value. 
This distinction motivates frameworks that weaken the constraints by including parameters which vary throughout the thermal history of the Universe, typically by means of phase transitions, see for example 
Refs.~\cite{Berkooz:2004kx,Bruggisser:2017lhc,Ellis:2019flb,Gan:2023wnp}.
In the context of Mesogenesis, the rate of $\cM\to\baryon\psi$ can be altered by a late-time phase transition which changes the effective Wilson coefficient controlling the decay~\cite{Elor:2024cea}. 
This, in turn, weakens the constraints, recovering regions of the Mesogenesis parameter space previously thought to be excluded.  

In this work, we propose a new Mesogenesis mechanism, dubbed \textit{Blocking Mesogenesis}, in which the mass of the dark fermion $\psi$ (and its decay products) are modified by a dark phase transition (PT) after $T_R$ and before BBN.
It is constructed such that the rate $\cM\to\baryon\psi$ is allowed before the PT and is kinematically blocked after it, see Fig.~\ref{fig:timeline}. 

The blocking Mesogenesis mechanism has several consequences.
(i)~$\BR(\cM\to\baryon\psi)$ can be as large as $\sim0.1$ in the early Universe and vanish today; thus, it can be mediated by new physics at sub-TeV scales, which can be probed directly by the LHC, rather than by rare meson decays. 
In addition, the required CP asymmetry can be smaller and even possibly compatible with SM sources of CPV in the case of $B^0$ and $B^0_s$ mixing. 
(ii)~The early Universe value of $m_\psi$ can be lighter than the proton without being excluded by proton decay bounds (which is not the case in Ref.~\cite{Elor:2024cea}).  
As such, the blocking mechanism allows for Mesogenesis models based on $D\to\baryon\psi$.  
Finally, we note that there exist ranges of $m_\psi$ for which the morphing successfully blocks the problematic channels while allowing for interesting signals in baryonic decay channels, e.g. $\Lambda_{b,c}\to \cM \MET$.

The rest of the paper is organized as follows.
In section~\ref{sec:explicitmodels} we present three mechanisms for blocking Mesogenesis. 
We provide models implementing each mechanism, and discuss their signals and constraints from terrestrial experiments. 
Our results are presented in section~\ref{sec:paramspace}. 
We provide examples of dark PTs which can accomplish the required mass morphing in section~\ref{sec:blocking}. 

\section{Benchmark models and terrestrial constraints}
\label{sec:explicitmodels}

\begin{figure}[t]
    \centering
    \begin{tikzpicture}[font=\large]
        \begin{feynman}
            \vertex (a1) {\(u\)};
            \vertex[right=1.5cm of a1] (a2);
            \vertex[right=1.5cm of a2] (a3);
            \vertex[right=1.5cm of a3] (a4);
            \vertex[right=1.5cm of a4] (a5){\(u\)};
            \vertex[below=0.75cm of a1] (b1){\(u\)};
            \vertex[below=0.75cm of b1] (c1);
            \vertex[below=0.45cm of c1] (d1) {\(d\)};
            \vertex[right=1.5cm of b1] (b2);
            \vertex[right=1.5cm of d1] (d2);
            \vertex[right=3.cm of c1] (c3);
            \vertex[right=1.5cm of c3] (c4);
            \vertex[above=0.75cm of c4] (b4);
            \vertex[below=0.75cm of a5] (c5){\(\bar{d}\)};
            \vertex[below=1.5cm of c5] (d4){\(\psi\)};
            \diagram*{
            (a1)--(a2)--(a3)--(a4)--(a5),
            (b2)--[boson, edge label'=\(W^{-}\)](d2),
            (b1)--(b2)--[edge label=\(s\)](c3),
            (d1)--(d2)--[edge label'=\(c\)](c3)--[scalar, edge label=\(Y^*\)](c4)--(c5),
            (c4)--(d4)
            };
            
            \draw[decoration={brace}, decorate]
              (d1.south west) -- (a1.north west)
              node[pos=0.5, left]{\Large\(p\,\)};
            
            \draw[decoration={brace,mirror}, decorate]
              ([xshift=0.2mm]c5.south east) -- ([xshift=0.2mm]a5.north east)
              node[pos=0.5, right]{\Large\(\,\pi\)};
        \end{feynman}
    \end{tikzpicture}
    \caption{\justifying The leading contribution to proton decay induced by Eq.~\eqref{eq:LprotalD}.}
    \label{fig:proton_decay}
\end{figure}
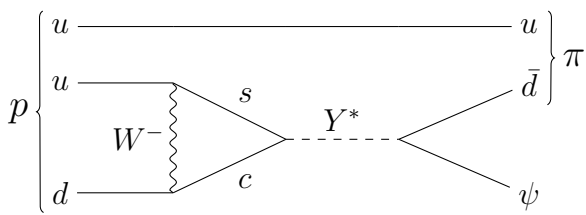

The ingredients for a successful model of blocking Mesogenesis are similar to those of the original Mesogenesis -- late-time abundance of $\cM \overline{\cM}$ followed by CP violating $\cM\to\baryon\psi$ decay-- with the addition of a dark phase transition that increases $m_\psi\to m_\psi+\Delta m_\psi$ such that $m_\psi+\Delta m_\psi>m_\cM-m_\baryon$ (the initial bound of $m_\psi>m_p$ is not relevant, see below). 
This requires that the dark phase transition be completed after $T_R$ and before BBN.
In this section, we introduce three mechanisms for blocking Mesogenesis, based on $B$ and $D$ decays:
\begin{itemize}
  \item[\textit{\textbf{A.}}] 
  $B_{d,s}^0$ decays with CPV in meson oscillations,
  \item[\textit{\textbf{B.}}] 
  $B$ decays with direct CPV, and
  \item[\textit{\textbf{C.}}] 
  $D$ decays with direct CPV.
\end{itemize}
The CPV in mechanism \textit{\textbf{A.}} can originate from the SM or from BSM sources. \textit{\textbf{B.}} and \textit{\textbf{C.}} necessarily require BSM sources of CPV.

Regarding Mesogenesis via $D^0$ meson oscillations; following~\cite{Nelson:2019fln, Alonso-Alvarez:2021qfd}, we solved the  Boltzmann equations to obtain the prediction for $\Ybar$.
Using the measured CP observable $[A^{D_{0}}_{\rm mix}]_{\rm exp}=(0.010\pm0.032)$ from~\cite{LHCb:2021ykz}, and assuming the maximal value of the mediator couplings, $y\sim \sqrt{4\pi}$ (see Sec.~\ref{sec:paramspace}), we find $\Ybar/\YBAU\lesssim 0.4$ for masses $m_\cY \gtrsim 200\eV[G]$, below which the characteristic signals from Mesogenesis models are well excluded. We conclude that $D^0$ meson oscillations is an unlikely Mesogenesis mechanism, and do not consider it further.
\paragraph*{Models:} To implement each of the three mechanisms, we construct two models, starting from the Mesogenesis models of~\cite{Elor:2018twp}. 
Both models introduce a new colored scalar mediator $\cY\sim(3,1)_{-1/3}$ with an $\cO(\TeV)$ mass, which acts as a \textit{portal} between the SM and a minimal dark sector -- which consists of the fermion $\psi$, its decay products, $\phi,\xi$, and a sector that facilitates the morphing.
Note that the present model differs from \emph{Mesogenesis with a Morphing Mediator} \cite{Elor:2024cea} in that now the dark sectors themselves are morphed rather than the heavy mediator $\mathcal{Y}$ . Schematically,
\begin{align}
    \label{eq:L}
    \cL 
    \supset 
    \cL_{\rm portal}+\cL_{\rm dark} + \cL_{\rm morph} \,,
\end{align}
where $\cL_{\rm dark}$ is responsible for the washout--preventing decay of $\psi$.
The portals for $B$ decays (mechanisms \textit{\textbf{A.}} and \textit{\textbf{B.}}) and $D$ decays (mechanism \textit{\textbf{C.}}) are given by,
\begin{align}
    \label{eq:LprotalB}
    \cL^B_{{\rm portal}} 
    = 
    &- y_{bc} \bar{c}_{R} b^{c}_{R} \cY^{*} - y_{s\psi} s^{c}_{R} \overline{\psi}_{R} \cY +\text{h.c.} \\
    \label{eq:LprotalD}
    \cL^D_{{\rm portal}} 
    = 
    &- y_{sc} \bar{c}_{R} s^{c}_{R} \cY^{*} - y_{d\psi} d^{c}_{R} \overline{\psi}_{R} \cY +\text{h.c.}
    \,,
\end{align}
in which the color indices are suppressed and are understood as totally anti-symmetrized for the $\cY q q'$ term and as $\delta_{a}^b$ contracted for the $\cY q \psi$ term~\cite{Alonso-Alvarez:2021qfd}. 
The first portal facilitates various $B$ meson decays, among which we will consider $B_d\to\Xi_c\psi$, $B_s\to\Omega_c\psi$. 
Including other channels (such as charged $B$ decays) relaxes the constraints on the couplings, and so we will not consider them here.  
The second portal allows for the $D$ meson decay $D \to \Lambda^0 \psi$. 

Mechanisms \textit{\textbf{B.}} and \textit{\textbf{C.}} involve direct CPV in $\cM\to\baryon\psi$ decays, which requires two amplitudes with different strong and weak phases. 
We leave the full model building for future work and here assume a priori direct CPV at the $\sim10^{-3}$ level. 
In addition, we assume there exist additional CP violating $\cM$ decay modes such that the $\cM$ and $\overline{\cM}$ total widths are equal -- as expected from the CPT theorem. 

Note that, in order not to wash out the BAU, these additional CP-violating modes should not violate baryon number, for example $\cM\to\bar{\psi}\psi+$SM particles.

As mentioned, we prevent washout by inducing prompt $\psi$ decay to a pair of stable particles: $\phi$, a SM-singlet scalar which carries Baryon number; and $\xi$, a SM-singlet Majorana fermion. Their interaction with $\psi$ is given by,
\begin{align}
    \label{eq:Ldark}
    \cL_{\rm dark} 
    &= 
    y_{\rm \psi\xi} \bar{\psi} \phi \xi + \text{h.c.}\,.
\end{align}
Both $\phi$ and $\xi$ are stabilized under a dark sector $\mathbb{Z}_2$ symmetry, which, along with the kinematic requirement that $|m_\phi - m_\xi| < m_p$, is necessary to prevent $\psi$-mediated decays back into the visible sector. Such a decay back into the visible sector which would wash out the generated BAU. 
As an added benefit, either can be candidates for dark matter~\cite{Elor:2018twp,Elahi:2021jia}.
Implementations of $\cL_{\rm morph}$ will be discussed in section~\ref{sec:blocking}.

\paragraph*{Proton decay}
Both portals from Eqs.~\eqref{eq:LprotalB}-\eqref{eq:LprotalD} induce proton decay for $m_\psi<m_p$.
An important consequence of the $m_\psi$ morphing between $T_{R}$ and BBN is a relaxation of the $m_\psi<m_p$ bound originating from proton stability. 
Assuming $m_p<m_\psi$ today, we need only show that the proton lifetime is significantly longer than the age of the Universe at BBN, $\sim$1 second.
We only evaluate the proton lifetime before the dark phase transition for the $D$ portal, Eq.~\eqref{eq:LprotalD}, since this is the only mechanism necessarily excluded by the proton lifetime bounds.
The leading diagram for proton decay is shown in Fig.~\ref{fig:proton_decay}.
We estimate the proton lifetime by integrating out both the $W$ boson and the new mediator $\cY$, resulting in a four-fermi $ \bar{c}_{R} b^{c}_{R} s^{c}_{R} \overline{\psi}_{R}$
interaction, which is loop-, CKM-, and chirally suppressed. The Wilson coefficient can be estimated as,
\begin{align}
    C_{\psi c\bar{d}s} 
    \sim 
    \frac{G_F}{8\sqrt{2} \pi^{2}} \frac{m_c m_s}{m_{\cY}^{2}} V_{cd} V_{us}  y_{cs} y_{d\psi} \,,
\end{align}
which, following~\cite{Nath:2006ut}, gives a proton lifetime of,
\begin{align}
    \tau_{p}
    \sim
    4\times10^{5}\left(\frac{4\pi}{y_{d\psi}y_{cs}}\right)^{2}\left(\frac{m_{\cY}}{400\eV[G]}\right)^{4}\text{sec}\, ,
\end{align}
where we conservatively assume $m_\psi=0$.
For values of $y_{d\psi}$, $y_{cs}$ and $m_\cY$ which can successfully generate the BAU (see Fig.~\ref{fig:D_dir_constraints}), the proton lifetime is longer than the age of the Universe at BBN~\cite{ParticleDataGroup:2024cfk}. 
\paragraph*{Terrestrial constraints}
Since the color-triplet mediator $\cY$ can have a mass of $\sim$TeV, it is accessible to collider experiments. 
In addition to its QCD interaction, it also has Yukawa-like interactions with $qq'$ and $q\psi$, see Eqs.~\eqref{eq:LprotalB}-\eqref{eq:LprotalD}.
These interactions lead to a variety of experimental signatures~\cite{Alonso-Alvarez:2021qfd} (see also~\cite{Hiller:2026osz}), which we summarize here and consider below. 
\paragraph{} 
LHC searches for jets or dijets accompanied by missing transverse energy~\cite{ATLAS:2017bfj,ATLAS:2017txd,ATLAS:2020syg,CMS:2019zmd,ATLAS:2024lpr} probe scenarios in which the mediator $\cY$ is produced in strong interactions and decays into a quark and dark-sector states. 
The resulting bounds on the corresponding branching fractions for $m_\cY \gtrsim 0.5\eV[T]$ were analyzed and recast in~\cite{Alonso-Alvarez:2021qfd}. 
We adopt these results to constrain the Yukawa couplings appearing in Eqs.~\eqref{eq:LprotalB} and~\eqref{eq:LprotalD}. For $m_\cY\lesssim0.5\,\TeV$ monojet $pp\to j\MET$ searches~\cite{ATLAS:2017bfj} strongly constrain the couplings $y_{q\psi}$ and $y_{q\psi}y_{qq'}$ via the processes $qq'\to \psi \psi +j$ and $qq'\to \psi q$, respectively. We generated the relevant LO cross sections using \textsc{MadGraph5}\_a\textsc{MC@NLO}~\cite{Alwall:2014hca}, with hadronization and showering implemented in Pythia8~\cite{Sjostrand:2014zea} and detector effects and signal cuts approximated by Delphes~\cite{deFavereau:2013fsa}. Constraints on the couplings were then obtained by comparison to the model-independent results in Table 6 of~\cite{ATLAS:2017bfj}.  
\paragraph{} 
Searches for four-jet final states at the LHC probe strong interaction pair production of mediators, each decaying into a pair of quarks. 
Constraints on the production cross section of color-triplet scalars decaying to diquark final states are presented in~\cite{ATLAS:2017jnp}, under the assumption $\Br{\tilde t \to qq'} = 1$, where $\tilde t$ denotes the color-triplet scalar. 
In our model, however, $\Br{\cY \to qq'} \neq 1$. 
We therefore reinterpret the results of~\cite{ATLAS:2017jnp} by the appropriate rescaling of the branching fractions.
Along with dijet resonance searches (see below), these provide the most stringent bounds on the couplings at low $m_\cY$, see Fig.~\ref{fig:B_mix_constraints} (right). 
\paragraph{} 
Dijet resonance searches are sensitive to the branching fraction of the mediator into diquark final states times the diquark coupling, $y_{qq'}$. 
We consider searches from ATLAS, CMS, LEP, and the TeVatron~\cite{CMS:2016ltu,ATLAS:2017eqx,CMS:2018mgb,ATLAS:2018qto, Dobrescu:2021vak,ATLAS:2025okg}. 
For several of these bounds~\cite{CMS:2018mgb,ATLAS:2017bfj}, we adopt the recasting of Ref.~\cite{Alonso-Alvarez:2021qfd}.
Refs.~\cite{Dobrescu:2021vak,ATLAS:2025okg} place upper limits on a color singlet resonance with a universal coupling to the quarks ($g_B$). 
We recast these limits on $g_B$ for the non-universal $y_{qq'}$ couplings of Eqs.~\eqref{eq:LprotalB}-\eqref{eq:LprotalD}, according to,
\begin{align}
    y_{qq'}^2 \Br{\cY \to qq'}
    =
   2 [g_B]_{\rm UL}^2 \frac{\sum_q L_{q\bar{q}}}{L_{qq'}} \,,
\end{align}
where $[g_B]_{\rm UL}$ denotes the 95\,\% upper limit on the $Z'$ coupling from the original analysis (assuming a universal quark coupling),
$L_{qq'}$ are the $qq'$ quark parton luminosity functions evaluated at the resonance mass and the sum runs over $q=u,d,s,c,b$ and the factor of 2 is from chirality.
\paragraph{} 
The dijet angular distributions constrain the $y_{qq'}$ couplings.
By recasting the TeVatron~\cite{D0:2009rxw,Dobrescu:2021vak}, we find $y_{qq'} \lesssim 3.8$, which is weaker than our conservative perturbative limit, $y_{qq'} < \sqrt{4\pi}$.

Finally, we emphasize that bounds from $B\to \baryon\MET$ are not relevant due to the $m_\psi$ morphing mechanism. 
 
\section{Results}
\label{sec:paramspace}

For each of the three mechanisms under consideration, we present the results of our analysis in the form of exclusion plots on the $m_\cY-m_\psi$ parameter space. 
For convenience, we define a reference $\cM\to\baryon\psi$ branching ratio value as
\begin{align}
    \widetilde{\BR}_{\cM\to\baryon\psi}^x
    \equiv
    \BR(\cM\to\baryon\psi) \, ,
\end{align}
evaluated at $m_\cY=400\, \eV[G]$ and for couplings $y_{qq'}^2y^2_{\tilde{q}\psi}=x$, which we use to illustrate benchmark values of the relevant parameters for which each mechanism can generate the full BAU.

\begin{figure*}[t]
    \begin{subfigure}{\textwidth}
    \centering
    \includegraphics[width=0.49\linewidth]{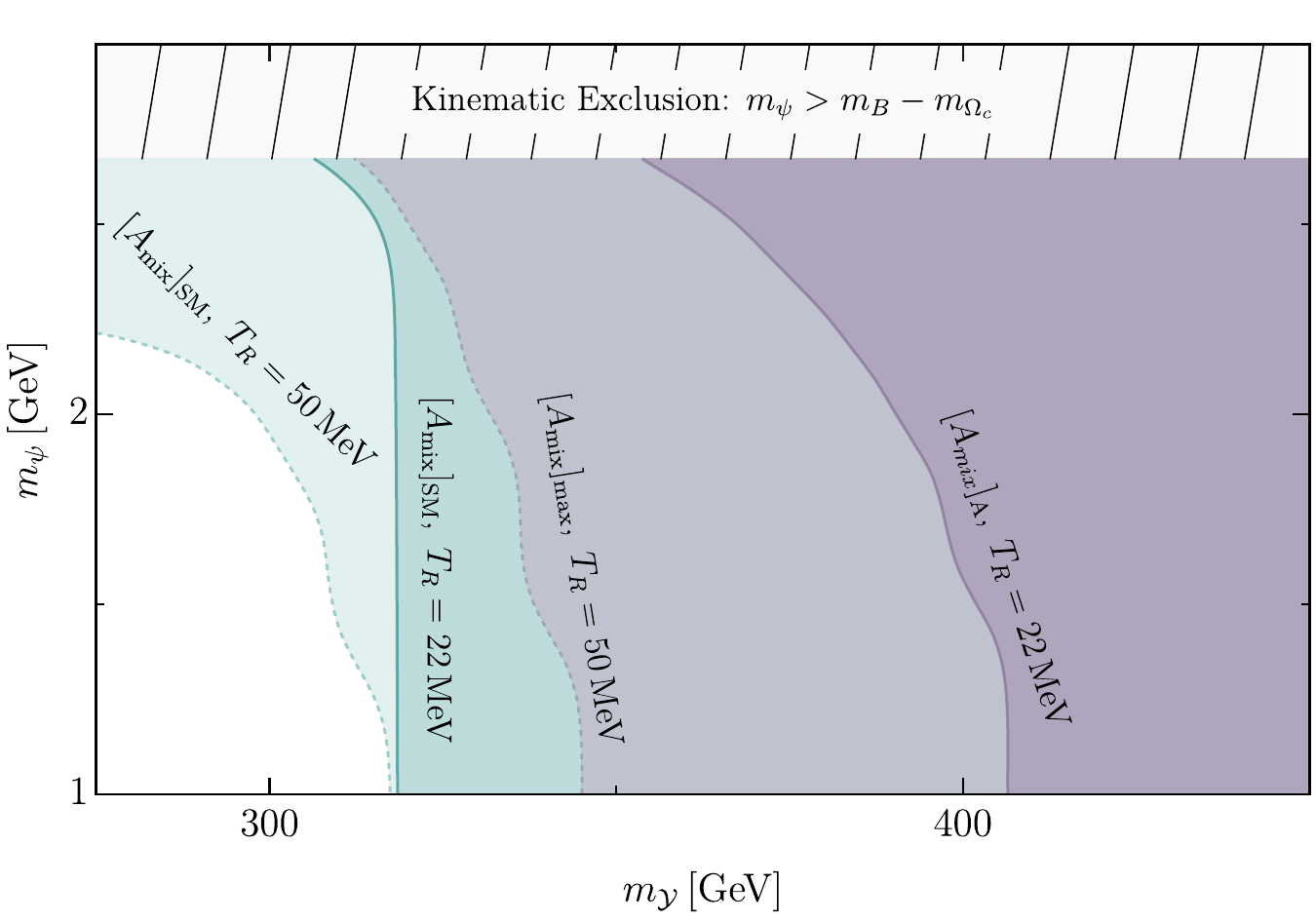}
    \includegraphics[width=0.49\linewidth]{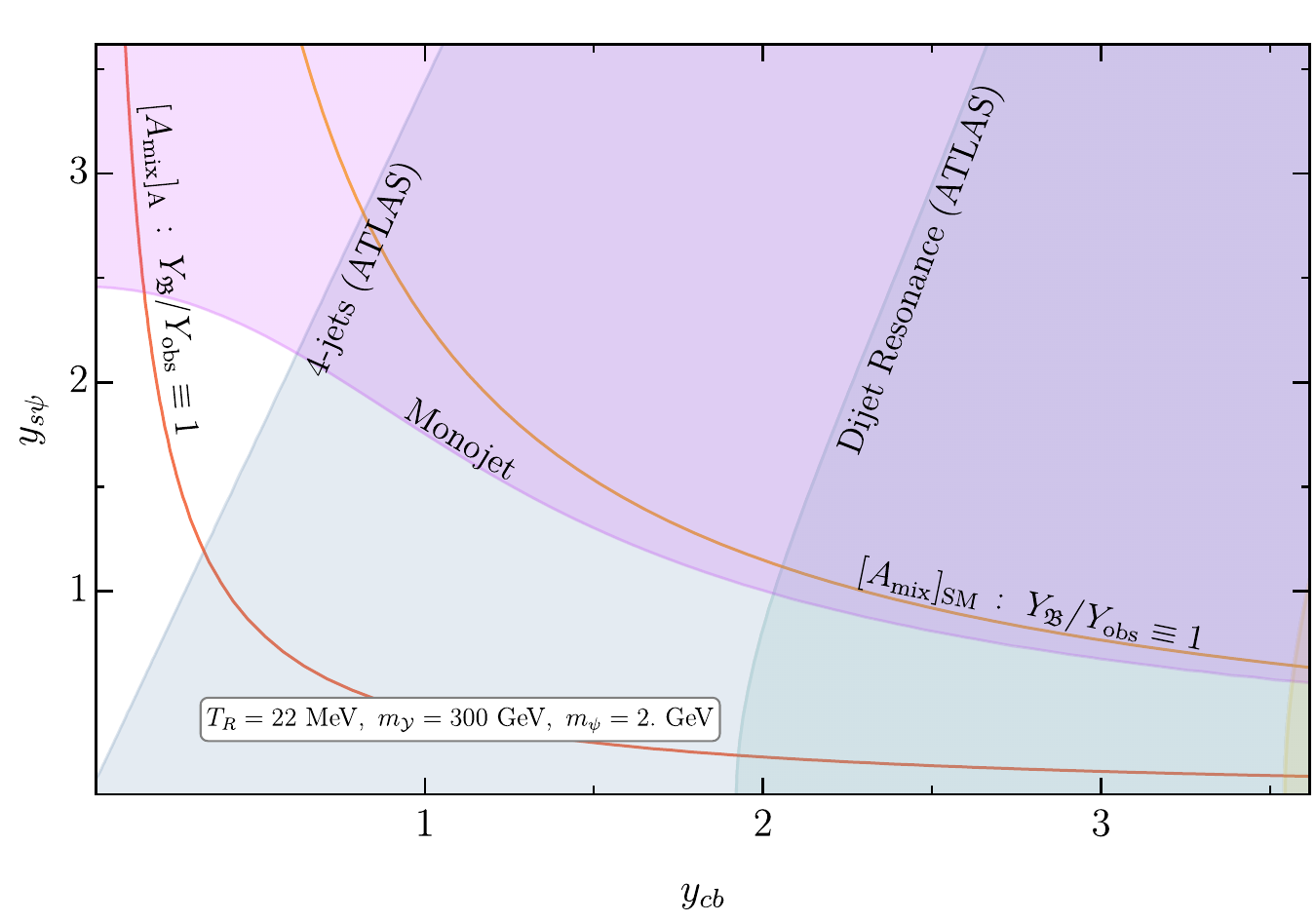}
    
    \caption{Mechanism \textbf{\textit{A}}.
    Values for asymmetries are given in the text.}
    \label{fig:B_mix_constraints}
\end{subfigure}
\begin{subfigure}{\textwidth}
    \centering
    \includegraphics[width=0.48\linewidth]{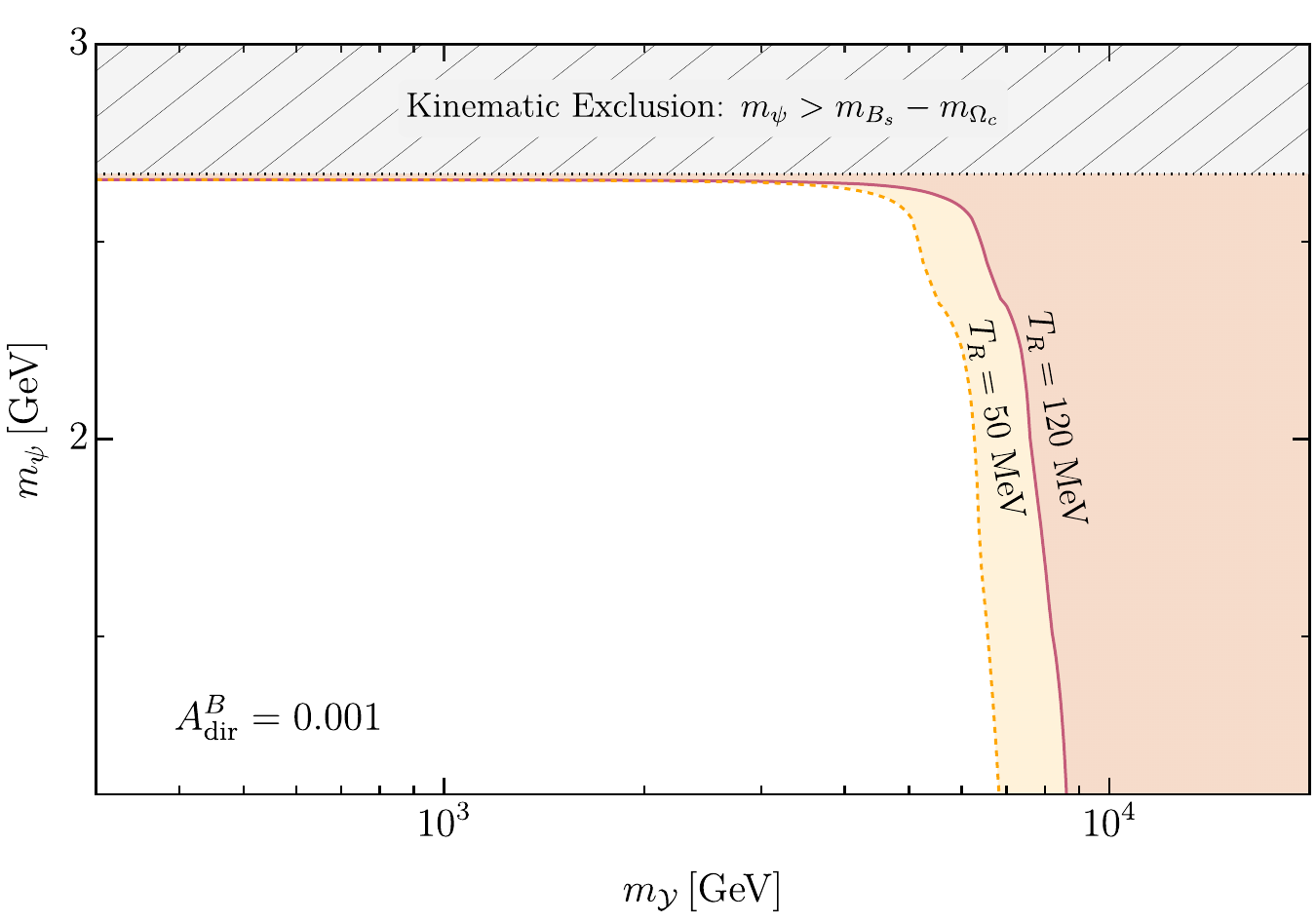}
    \includegraphics[width=0.5\linewidth]{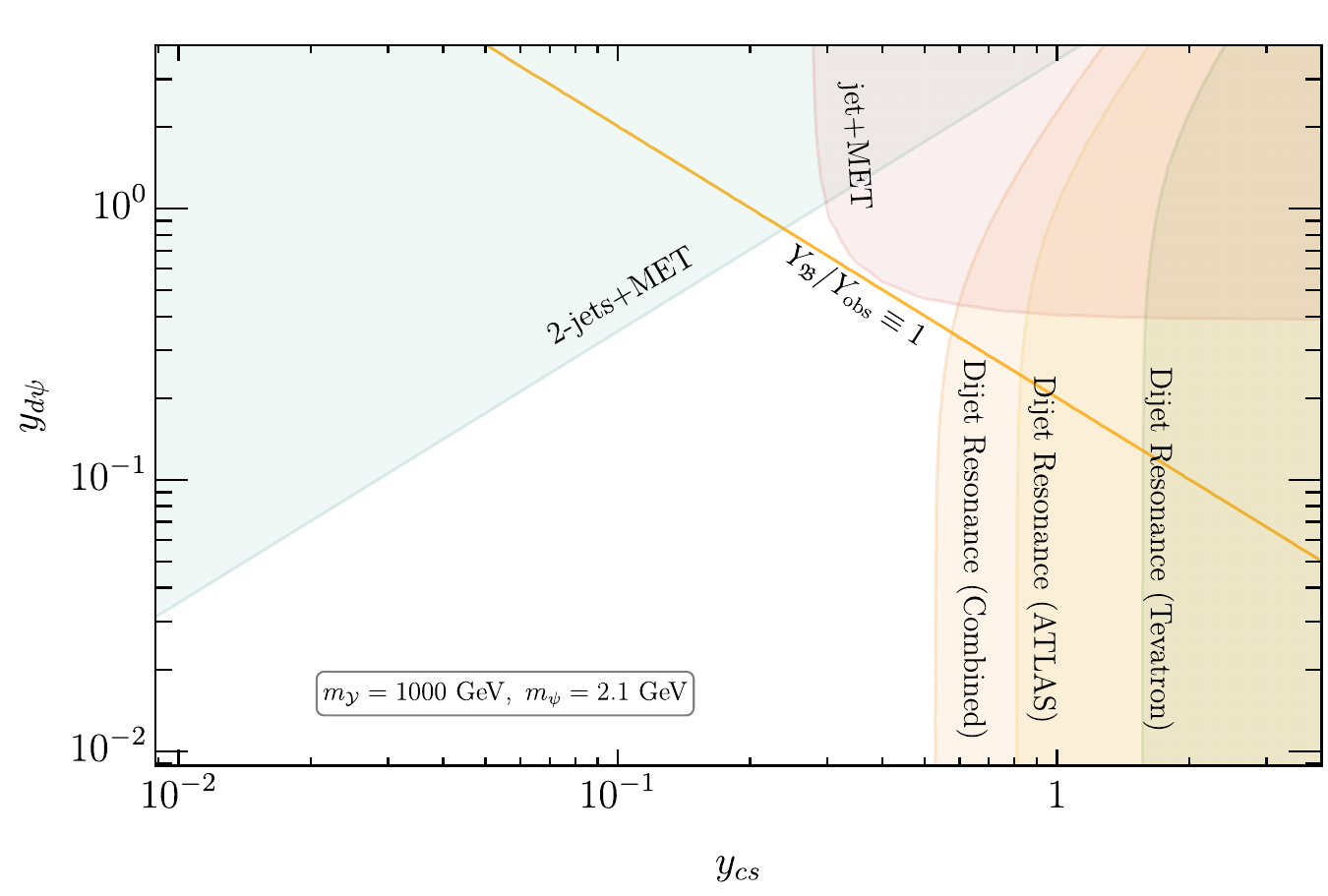}
    \caption{
    Mechanism \textbf{\textit{B.}} with $\ACPdir=0.001$.}
    \label{fig:B_dir_constraints}
\end{subfigure}
\begin{subfigure}{\textwidth}
    \centering
    \includegraphics[width=0.49\textwidth]{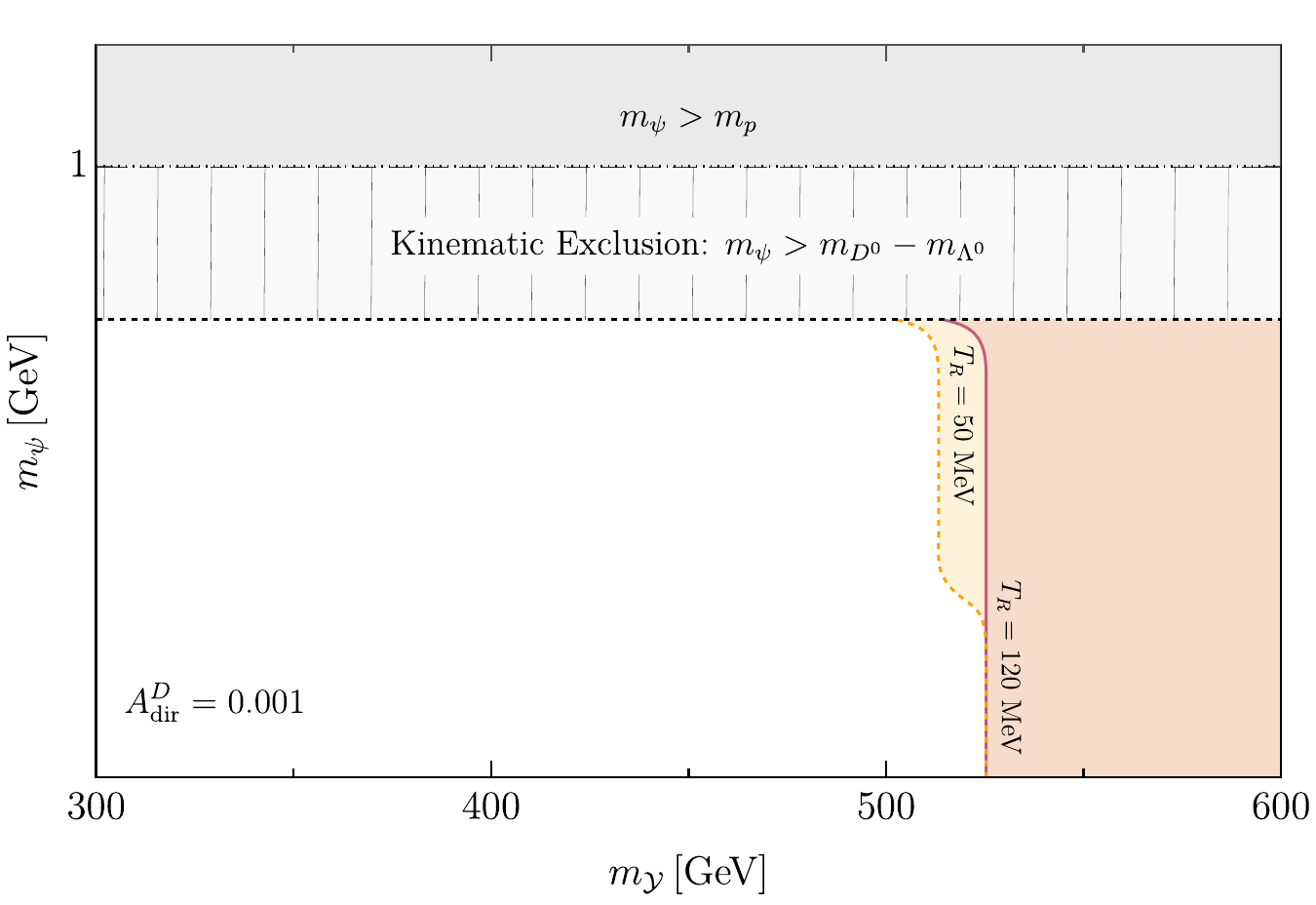}
    \includegraphics[width=0.5\textwidth]{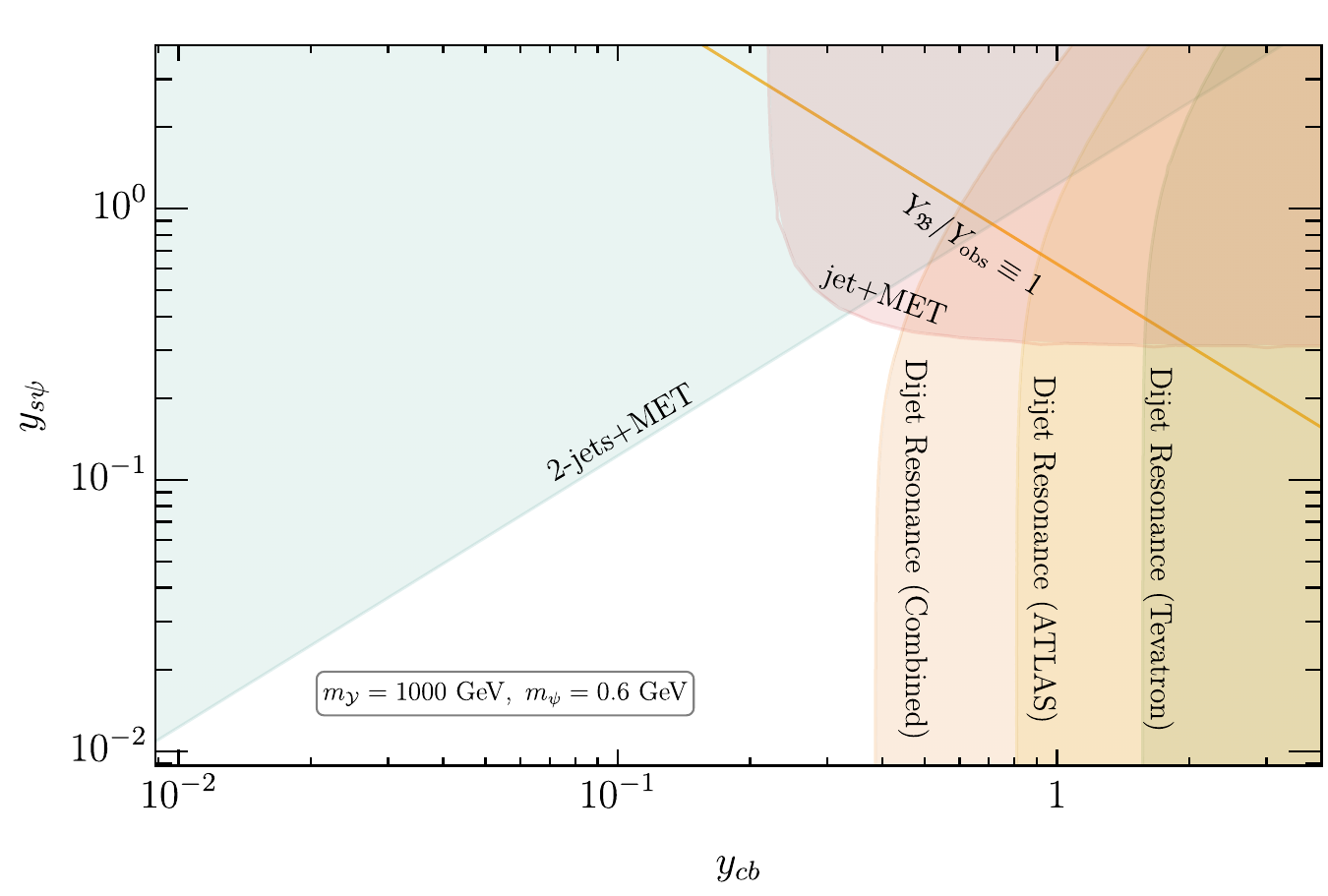}
    \caption{
    Mechanism \textbf{\textit{C.}} with $\ACPdir=0.001$. }
    \label{fig:D_dir_constraints}
    \end{subfigure}
    \caption{\justifying 
    (left) Regions excluded by the combination of terrestrial constraints and BAU generation for various values of the reheating temperature $T_R$ in each of the mechanisms discussed in the text. 
    (right) For a sample point in the $m_\cY-m_\psi$ parameter space, we show how the couplings are constrained by terrestrial experiments and BAU generation.}
\end{figure*}

\paragraph*{\textbf{A.}} For this mechanism, since both $B_s$ and $B_d$ mesons are produced in the early universe, the decays of both mesons contribute to BAU generation. 
From Eq.~\eqref{eq:YBmeso} we obtain, 
\begin{align}
     \frac{\Ybar^{\rm mix}}{\YBAU} 
     \simeq &
    \frac{y_{bc}^2 y_{s\psi}^2}{0.25}
    \!\left(\frac{400 \eV[G]}{m_\cY}\right)^4 
    \!\!\!\times \left( 1.01F_s-0.01 F_d \right)\ ,
\end{align}
where,
\begin{align}
    F_q 
    \equiv
\frac{\widetilde{\BR}^{0.25}_{B_q^0\to\baryon\psi} }{\eta_q\times 10^{-3}} 
    \frac{A^{B_q}_{\rm mix}}{[A^{B_q}_{\rm mix}]_{\rm A}}   \frac{\alpha^{\rm mix}_{B_q}(T_R)}{\alpha^{\rm mix}_{B_q}(22\,\eV[M])} \, ,
\end{align}
with $\baryon=\Xi_c,\,\Omega_c$ $\eta_q = 3,4$ for $q=d,\,s$, respectively.   
The $B^0_{d,s}$ decay rates are taken from Ref.~\cite{Elor:2022jxy}.
The CP asymmetries in $B$ meson mixing are given by $[A^{B_d}_{\rm mix}]_{\rm SM}=(-4.7\pm 0.4)\times 10^{-4}$ and $[A^{B_s}_{\rm mix}]_{\rm SM}=(2.1\pm0.2)\times 10^{-5}$~\cite{Lenz:2019lvd} with corresponding upper limits ($2\sigma$) of $|[A^{B_d}_{\rm mix}]_{\rm UL}|<5.5\times 10^{-3}$ and $|[A^{B_s}_{\rm mix}]_{\rm UL}|<5\times 10^{-3}$~\cite{ParticleDataGroup:2024cfk,HeavyFlavorAveragingGroupHFLAV:2024ctg}, respectively. 
However, as shown in~\cite{Miro:2024fid}, NP contributions to CP asymmetries cannot saturate these bounds.
We therefore consider the upper limits estimated in~\cite{Miro:2024fid}, $[A^{B_d}_{\rm mix}]_{\rm A}=-4\times 10^{-4}$ and $[A^{B_s}_{\rm mix}]_{\rm A}=5\times 10^{-4}$ as our benchmark values.
The functions $\alpha^{\rm mix}_{B_{d,s}}(T_R)$ are taken from~\cite{Alonso-Alvarez:2021qfd}.
Fig.~\ref{fig:B_mix_constraints} (left) shows regions of the $m_\cY-m_\psi$ plane excluded by the combination of terrestrial constraints from Section~\ref{sec:explicitmodels} and the requirement $\Ybar/\YBAU \approx 1$. 
Different regions are obtained for different values of the reheating temperature and the CP asymmetry.  
The sea-green regions exclude mechanisms where the central value of the SM prediction is taken for the CP asymmetry, with the solid and dashed lines corresponding to two different temperatures, and the purple regions correspond to the maximal allowed CP asymmetry based on the analysis of~\cite{Miro:2024fid}.  
In the right panel of Fig.~\ref{fig:B_mix_constraints}, we show the individual bounds in the $y_{cb}-y_{s\psi}$ plane for benchmark values of the masses and the temperature.
This result suggests that blocking Mesogenesis with SM CPV alone is strongly under tension. 
However, sub-TeV signals are not excluded within the upper limit of observed CPV in $B$-meson mixing. 
\paragraph*{\textbf{B.}} In this mechanism, we consider neutral $B$ decays with direct CPV from the dark sector. 
Since the CPV is direct, we consider only one decay channel, and note that the inclusion of the other decays induced by Eq.~\eqref{eq:LprotalB} only relaxes the constraints on the couplings. 
We find,
\begin{align}
    \frac{\Ybar^{\rm dir}}{\YBAU} 
    \simeq& 
     \left(\frac{y_{bc}^2 y_{d\psi}^2}{0.04}\right) \left(\frac{1000\eV[G]}{m_\cY}\right)^4 
     \left(\frac{\widetilde{\BR}^{0.04}_{B_s\to\Omega_c\psi}}{6\times10^{-6}}\right)
     \nonumber \\
    &\times 
     \left(\frac{A^{B}_{\rm dir}}{10^{-3}}\right)  \left(\frac{\alpha^{\rm dir}_{B}(T_R)}{\alpha^{\rm dir}_{B}(100\, \MeV)}\right)\,,
    \label{eq:BdirYrat}
\end{align}
where the $\alpha$ function for direct CPV is given by~\cite{Elor:2025fcp},
\begin{align}
    \alpha^{\rm dir}_\cM(T_R) 
    \approx
   220\frac{T_R}{100\,\eV[M]}\frac{2m_\cM}{m_\Phi}\ .
\end{align}
The results from this mechanism are presented in Fig.~\ref{fig:B_dir_constraints}, where we consider $A^B_{\rm dir}=10^{-3}$ and $T_R=50\,\MeV$ and $120\,\MeV$. 

Inspired by~\cite{Hiller:2026osz}, for $2.7\eV[G]\lesssim m_\psi+\Delta m_\psi \lesssim 4.1\eV[G]$, this model admits a signal in the form of the decay channel $\Xi_b^0\to D^0\psi$, which appears in experimental searches as $\Xi_b^0\to D^0+\MET$. 
\paragraph*{\textbf{C.}} 
To the best of our knowledge,
unlike $B\to \baryon \psi$, there are no theoretical predictions for the branching fraction of the $D\to\baryon\psi$ decay modes. 
Therefore, we estimate the relevant branching fractions from the $c\to s s \psi$ inclusive rate with an appropriate change of the phase space as in Ref.~\cite{Alonso-Alvarez:2021qfd} for $b$ decays. 
We estimate,
\begin{align}
    \label{eq:Destimate}
    \BR(D^0\to \Lambda\psi )
    &=
    \tau_{D^0}\frac{\gamma(m_{\Lambda})}{ \gamma(\delta m)}
    \frac{y_{cs}^2y_{d\psi}^2}{m_\cY^4}
    \frac{m_c^5}{4608\pi^3 }g\left(\frac{m_\psi^2}{m_c^2}\right)\ , 
\end{align}
where $\delta m=m_c-m\psi$, $m_\Lambda=1.115\eV[G]$,  $\gamma(m)$ is defined in section IV.D of~\cite{Alonso-Alvarez:2021qfd} and $g(x)=-1-8x +8x^3 - x^4 -12 x^2 \ln{x}$. 
Multiplying by the ratio $\gamma(m_\Lambda)/\gamma(\delta m)$ simply corresponds to cutting off the phase space integration in the inclusive decay rate at $m_{\Lambda^0}$.
As a cross-check, we compared our rate to an estimation based on the treatment of Ref.~\cite{Hou:2005iu} and find agreement up to a factor of $\sim 2$.

Combining Eqs.~\eqref{eq:YBmeso} and~\eqref{eq:Destimate}, we find 
\begin{align}
    \frac{\Ybar^{\rm dir}}{\YBAU} 
    \simeq &
    \left(\frac{y_{sc}^2 y_{s\psi}^2}{10^{-2}}\right) \left(\frac{400 \eV[G]}{m_\cY}\right)^4 
    \left(\frac{\widetilde{\BR}^{0.01}_{D^0\to\Lambda\psi}(m_\psi)}{2\times 10^{-5}}\right)
    \nonumber \\
    &\times 
      \left(\frac{A^{D}_{\rm dir}}{10^{-3}}\right) \left(\frac{\alpha^{\rm dir}_{D^0}(T_R)}{\alpha^{\rm dir}_{D^0}(100\, \MeV)}\right) \,.
     \label{eq:DdirYrat}
\end{align}
The allowed parameter space of this mechanism can be found in Fig.~\ref{fig:D_dir_constraints}. 

For $1\eV[G]< m_\psi + \Delta m_\psi < m_{\Lambda_c}-m_K\approx  1.7\eV[G]$,  this model admits a signal $\Lambda_c\to K \psi$ which appears in experimental searches as $\Lambda_c\to K \MET$~\cite{Hiller:2026osz}.
This channel was discussed in~\cite{Hiller:2026osz}, from which the branching fraction can be estimated as,
\begin{align}
    \BR\left(\Lambda_c \to K \psi\right) 
    \approx 
    4\times 10^{-8} \frac{y^2_{cs}y^2_{d\psi}}{10^{-2}}\left(\frac{m_\cY}{400\eV[G]}\right)^4\ , 
\end{align}
with an $\cO(1)$ dependence on $m_\psi$.
%
%

\section{The mass morphing}
\label{sec:blocking}

In this section, we discuss the morphing of the $\psi$ and $\phi$ masses required to kinematically block $\cM\to\baryon\psi$/proton decay. 
The temperature window for the morphing mechanism is between $T_R$ and before BBN, as shown in Fig.~\ref{fig:timeline}, where the reheating temperature is taken to be  $T_R\sim20\,\eV[M]\,(100\,\eV[M])$ for mechanism \textit{\textbf{A.}}\,(\textit{\textbf{B.}} and \textit{\textbf{C.}}). 
Schematically, we write,
\begin{align}
    m_i \to m_i + \Delta m_i\,,
\end{align}
for $i\in \{\psi,\phi\}$ (in principle, the mass of $\xi$ can also be morphed, although we do not consider this scenario in this work), from which the conditions for the kinematic blocking are,
\begin{align}
    \Delta m_\psi &> m_p - m_\psi\,, \\
    \Delta m_\psi &> m_\cM - m_\baryon  \,,
\end{align}
for proton decay and $B$ decay, respectively. 
To avoid washout of the asymmetry, the condition $|m_\phi+\Delta m_\phi-m_\xi|<m_p$ must still hold after morphing, which puts an upper limit on $\Delta m_\phi$ set by the choices of initial masses. This condition can be relaxed by morphing $\xi$ as well.
Additionally, \cite{Nelson:2019fln} showed that astrophysical constraints require $m_\phi > 1.2\eV[G]$ today, and as such the morphing of $\phi$ must accommodate this.

In the following, we introduce two realizations of dark-sector phase transitions (PTs) that generate the required $\Delta m_{\psi,\phi}$:
(i)~a strongly-interacting sector with a confining PT, and
(ii)~a dark Abelian Higgs model with a first order PT. 

\subsection{Strongly interacting dark sector}

First, we model the ``morphing'' sector with a strong non-Abelian interaction which confines at some temperature $T_c$. 
In this model, $\cL_{\rm morph}$ contains a scalar operator $\cO_D$ coupled to both $\psi, \phi$ through $\bar\psi\psi\cO_D$, $\phi^\dagger\phi\cO_D$. 
Below the confinement temperature, $T<T_c$, $\cO_D$ acquires a VEV $v_D$, inducing the mass shift $\Delta m_\psi = \bar{\psi}\psi \langle \cO\rangle$ (and similarly for $\Delta m_\phi$). 

As an explicit example, we consider $\cO_D=\bar{\Psi}_{D}\Psi_{D}$, where $\Psi_{D}$ transforms as the fundamental of some dark SU($N_s$). 
To avoid constraints form BBN, we assume and light states in the morphing sector are unstable.
The EFT Lagrangian includes:
\begin{align}
    \cL_{\rm morph}
    \subset
    \frac{g^{2}_\psi}{\Lambda^{2}}\bar{\psi}\psi\bar{\Psi}_{D}\Psi_{D}
    +
    \frac{g^{2}_\phi}{\Lambda}\phi^\dagger\phi\bar{\Psi}_{D}\Psi_{D},
\end{align}
where $\Lambda$ is the cutoff scale of the EFT and $g_\psi$ and $g_\phi$ are both dimensionless. 
At $T_c$, the bilinear acquires a VEV $\langle \bar{\Psi}_{D} \Psi_{D} \rangle =v_D^{3}$ where $v_D\gtrsim T_c$.
This leads to the following mass morphing,
\begin{align}
    \label{eq:mass_from_condesation}
    \Delta m_\psi 
    &\sim 
    1.8 \eV[G] \frac{g^{2}_\psi}{16\pi^{2}}\left(\frac{v_D}{45 \eV[M]}\right)^{3}\left(\frac{90\eV[M]}{\Lambda}\right)^{2}\, ,\\
    \Delta m_\phi^2
    &\sim (0.5\,\eV[G])^2\frac{g^{2}_\phi}{16\pi^{2}}\left(\frac{v_D}{45 \eV[M]}\right)^{3}\left(\frac{90\eV[M]}{\Lambda}\right)\ ,
\end{align}
where $\Lambda>v_D$ to ensure EFT validity. 
To avoid constraints from neutron stars~\cite{McKeen:2018xwc,Nelson:2019fln}, and prevent washout of the generated BAU (as mentioned in the previous section), we require $1.2\eV[G]<m_\phi+\Delta m_\phi<m_p+m_\xi$ after morphing. 
The range of allowed initial masses is $1.2\eV[G]-\Delta m_\phi<m_\phi\lesssim m_p+ m_\xi-\Delta m_\phi$. 
For example, for the benchmark values taken in Eq.~\eqref{eq:mass_from_condesation}, $m_\xi >0.3\eV[G]$ is necessary for this range to be well defined, with the upper limit $m_\xi < m_\psi - m_\phi$ needed to ensure prompt $\psi$ decay to the stable dark sector states. 
For mechanism \textbf{\textit{C.}}, since $m_\psi\lesssim 0.6 \eV[G]$, these benchmark parameters are not relevant, and a larger VEV is required (unless $\xi$ is also morphed). 
Regardless, we verified that for any legal choice of the initial masses given these constraints and for $y_{\psi\xi}\sim \cO(1)$, the $\psi$ lifetime is $\cO(10^{-23}s)$, far below the Hubble time at the relevant temperatures, meaning the decay $\psi\to \phi \xi$ is prompt as expected. 
\subsection{Dark Abelian Higgs}

\begin{figure}[t]
    \centering
    \includegraphics[width=\linewidth]{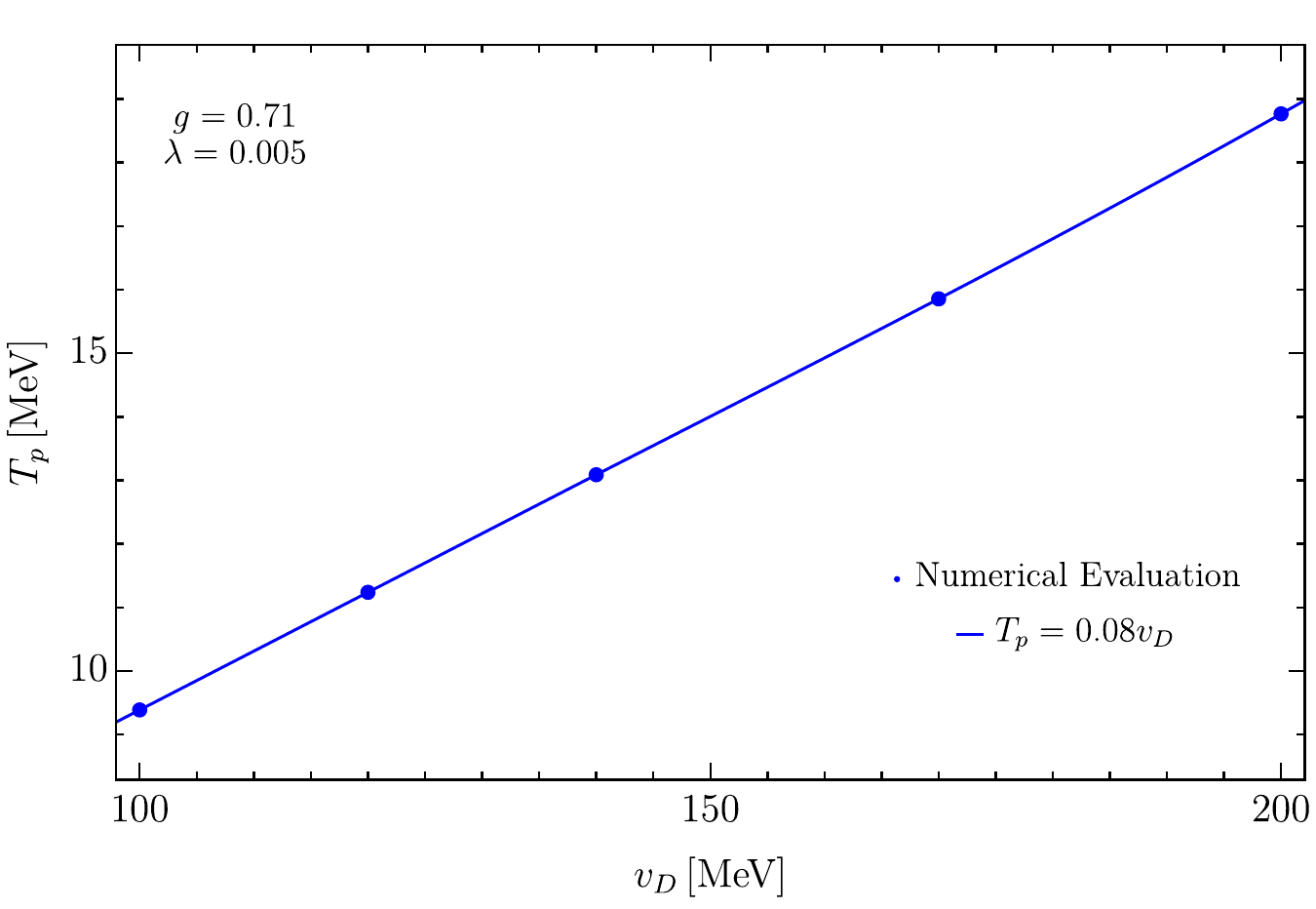}
    \caption{\justifying
    The percolation temperature as a function of the VEV for $g=0.71,\ \lambda = 0.005,\ y= \tilde \lambda =1$. We display mainly the values of $T_p$ relevant for mechanism \textbf{\textit{A.}}, and note that the relation remains linear for higher temperatures/larger VEVs}
    \label{fig:Tp}
\end{figure}

As a second example, following~\cite{Breitbach:2018ddu}, we consider the spontaneous breaking of a U(1)$_D$ gauge symmetry by a dark Abelian Higgs $H_D$ with unit U(1)$_D$ charge.
The scalar potential allows for a first order phase transition of the U(1)$_D$ symmetry, beginning at some critical temperature $T_c$ and proceeding via bubble nucleation until completion.
We note that this model does allow for continuous transitions, but that these cannot produce the required hierarchy between $T_p$ and $v$.
We consider the phase transition completed at the percolation temperature, $T_p$~\cite{Breitbach:2018ddu,Athron:2022mmm, Croon:2023zay}.
The interactions in $\cL_{\rm portal}^{B,D}$ (Eqs.~\eqref{eq:LprotalB}, \eqref{eq:LprotalD}) require $\psi$ to be a $U(1)_D$ singlet and therefore a $H_D\bar{\psi}\psi$ term is forbidden. 
Since such a term is necessary to induce the required mass shift, we introduce another vector-like fermion $\psi_2$ with a $U(1)_D$ charge, and generate the mass-shifting term through mixing of the two fermions. 
Overall, the morphing Lagrangian is given by 
\begin{align}
    \cL_{\rm morph} 
    &= 
    \cL_{\rm kin.} + V_{0}\left(H_{D},\phi\right) +\cL_\psi \, ,
\end{align}
where $\cL_{\rm kin.}$ contains all kinetic terms. 
The scalar potential is given by, 
\begin{align}
    \label{eq:V0}
    V_{0}\left(H_{D},\phi\right)
    =&
    -\frac{1}{2}\mu^{2} |H_{D}|^{2}  +\frac{\lambda}{4}|H_{D}|^{4} +\Tilde{\lambda}|H_{D}|^{2}\phi^{2} \nonumber \\
    &+ \frac{\lambda_\phi}{4} \phi^4 +\frac{1}{2}m_\phi^2 \phi^2 \, ;
\end{align}
the fermion mass and Yukawa terms are
\begin{align}
    \cL_\psi
    =
    -m_\psi\bar{\psi}\psi-m_2\bar{\psi}_2\psi_2
    -y_\psi\bar{\psi}\psi_2H_D+h.c. \, .
\end{align}
After $H_D$ acquires a VEV, the fermions mix, and the new mass eigenstates have,
\begin{align}
    m_{\pm} 
    = 
    \frac{m_\psi+m_2}{2}\pm \frac{1}{2}\sqrt{(m_2-m_\psi)^2-4y_\psi^2v_D^2}\,,
\end{align}
where we took $m_2 > m_\psi$. 
As required, the couplings of both $\psi$ and $\phi$ to the morphon induce the following mass changes, 
\begin{align}
    \Delta m_\phi^2 &= 2\tilde \lambda v_D^2\quad \\
    \Delta m_\psi &= \frac{m_2-m_1}{2} - \frac{1}{2}\sqrt{(m_2-m_1)^2-4y^2v_D^2}\,.
\end{align}
Details of our perturbative treatment of thermal effects in this model can be found in App.~\ref{app:Thermal}. 
For our analysis, we assumed the dark sector and SM were not in thermal equilibrium during the morphing, with $T_{\rm dark}/T_{\rm SM} \approx 0.5$~\cite{Breitbach:2018ddu}. 
As such, all temperatures relevant to the phase transition refer to the temperature in the dark sector. 
We compute the percolation temperature numerically using both the PT2GW Mathematica package~\cite{Brdar:2025gyo} and CosmoTransitions~\cite{Wainwright:2011kj}. 
Our results show that the percolation temperature is linear in the VEV, and we find that there exists a region of the parameter space $(\lambda,\ v_D)$ for which the percolation temperature is in the required range, see Fig.~\ref{fig:Tp}.
We conclude that this type of phase transition can reasonably achieve $\Delta m_\psi\sim \cO(100\eV[M])$ for mixing, and $\cO(\GeV)$ for direct CPV. 
For mechanism \textbf{\textit{A.}}, this means that this model is only relevant for near-threshold values of the initial $\psi$ mass. 
For the other two mechanisms, however, the full parameter space is comfortably covered due to the higher percolation temperatures. 

A general characteristic of FOPTs is the release of latent heat into the environment. 
In the case of cosmological FOPTs, the latent heat manifests as vacuum energy, which leads to three main effects; 
(i)~bubble-wall friction, in which the masses of particles coupled to the morphon increase to the true-vacuum values upon crossing the expanding bubble walls~\cite{Cohen:1990py,Baker:2019ndr}; 
(ii)~the production of gravitational waves through bubble collisions, plasma effects, etc.~\cite{Christensen:2018iqi}; 
(iii)~reheating of the dark sector thermal bath. 
We used~\cite{Brdar:2025gyo} to compute the normalized latent heat density and found $(\Delta V - \Delta \partial V|_{T=T_p})/\rho^*(T) \sim 0.1-0.3$
for our benchmarks.
We have verified that (i)~there is enough latent heat to modify the masses of all morphon-coupled particles present in the Universe during the PT; 
(ii)~that the energy released into gravitational waves is below observational limits (for verifying this we turned to~\cite{Brdar:2025gyo}), 
and (iii)~that the reheating of the dark sector is mild and does not interfere with $N_{\rm eff}$ constraints at BBN nor lead to a period of inflation. 

It is possible to obtain the necessary scale separation between the temperatures and mass differences in a supercooled phase transition, \ie~a first order phase transition in which the nucleation temperature is largely separated from the critical temperature, see for example~\cite{Baldes:2021aph,Athron:2022mmm,Li:2025nja}. However, such phase transitions necessarily lead to a period of cosmic inflation, washing out the generated BAU. Therefore, supercooled PTs may be relevant for Mesogenesis models in which the BAU is significantly overproduced.

\section{Conclusions}

In this work we introduced a novel class of Mesogenesis mechanisms in which the masses of the dark fermion, $\psi$, (and its daughter particles) are morphed between reheating and BBN. 
As a result, bounds on the Mesogenesis parameter space from rare meson decay into baryon and missing energy, $\BR(B\to\baryon\MET)\lesssim10^{-5}$, as well as bounds from proton stability, are avoided. 
The former allows for successful BAU generation with smaller CPV -- which in an extreme case can be based on SM CPV only in $B$-mixing. 
The latter restores the possibility of using $D\to \baryon\psi$ decays, which are na\"ively excluded by the proton stability bound on $m_\psi$. 

We demonstrated this mechanism for $B$ and $D$ decays with CPV in mixing and from a direct new physics source in the decay, and present the allowed parameter space for each benchmark model.  
We provided explicit examples for the dark-sector phase transition mechanism, and for each showed that the relevant morphing of the $\psi$ mass can be achieved. 
Terrestrial signals for the blocking mechanism are sub-TeV colored scalars, as well as rare decays of $\Lambda_c$ or $\Xi_c$ to lighter mesons and missing energy, for which future colliders such as FCCee may search.
If the dark-sector phase transition is first order, an additional potential signal comes in the form of gravitational waves. GW signals of the U(1)$_D$ model presented here have been extensively studied in the literature, see for example~\cite{Salvio:2023blb,Breitbach:2018ddu,Borah:2021ocu,Balan:2025uke,Feng:2026iqo}.

\section*{Acknowledgments}

We thank Miguel Escudero and Joachim Kopp for useful discussions and comments on the manuscript. 
We thank  Yann Gouttenoire, Yuval Grossman, Enrique Kajomovitz, Graham Kribs, Gilad Perez, Arttu Rajantie, Geraldine Servant, Yael Shadmi and Daiki Ueda for useful discussions.
This work is supported by the ISF (grant No. 597/24) and by BSF (grant No. 2024091). 
YS thanks CERN-TH for the scientific associateship.

\appendix

\section{Thermal theory setup for phase transition}
\label{app:Thermal}

In this appendix we provide additional details of the morphing sector of the model, specifically on our perturbative treatment of thermal effects.   
For reviews of Thermal Field Theory, see e.g.~\cite{Quiros:1999jp,Laine:2016hma}. Since the relevant temperature for our phase transition is $\cO(\MeV)$, and $\psi,\phi$ have $\cO(\GeV)$ masses, their thermal contributions are negligible. 
The effective tree-level potential is 
\begin{align}
    V_0 = -\frac{1}{2} \mu^2 h_D^2 + \frac{1}{4} \lambda h_D^4\,,
\end{align}
where $h_D$ is the radial model of the complex scalar $H_D$ that acquires a VEV. The field-dependent masses are,
\begin{align}
    m_h^2 &= \pdv[2]{V_0}{h} =-\mu^2 + 3 \lambda h^2\,, \\
    m_G^2 &= \frac{1}{h} \pdv{V_0}{h} = -\mu^2 + \lambda h^2\,, \\
    m_A &= g^2 h^2\,,
\end{align}
where $m_G$ and $m_A$ are the Goldstone boson and gauge boson masses, respectively, and $g$ is the gauge coupling. Thermal contributions from the light scalar $h_{D}$, the Goldstone boson and the gauge boson $A_{\mu}$ generate 1-loop thermal corrections to the potential,
\begin{align}
    V_{1T} &= \frac{T^4}{2\pi^2} \left(J_B\left(\frac{m_h^2}{T^2}\right) + J_B\left(\frac{m_G^2}{T^2}\right) + 3 J_B\left(\frac{m_A^2}{T^2}\right)\right)\,,
\end{align}
where,
\begin{align}
    \label{eq:J}
    J_{B}(y) =  \int_0^\infty \dd x x^2 \ln\left[1-e^{-\sqrt{x^2+y^2}}\right]\,.
\end{align}
We evaluate these expressions numerically without approximation.
Note it is necessary to count both the Goldstone mode and an un-physical thermal longitudinal mode of the gauge boson, see~\cite{Arnold:1992rz} for details. 
This unphysical mode will be resummed in the IR, along with the other divergent behaviour of the thermal 1-loop potential. This resummation is implemented by the inclusion of a ``daisy term"~\cite{Arnold:1992rz,PhysRevD.45.4695, PhysRevD.45.2933},
\begin{align}
    V_{\rm daisy} = -\frac{T}{12\pi} \sum_{i=h,G,L} \left[(m_i + \Pi_i)^3 - m_i^3\right]\,,
\end{align}
where $\Pi_i$ are the Debye masses of  $h_D$, the Goldstone and the \textit{longitudinal} polarization of the gauge boson (the transverse modes are not resummed),
\begin{align}
    \Pi_h &= \left(\frac{\lambda}{3}  + \frac{g^2}{4}\right) T^2 \\
    \Pi_G &= \left(\frac{\lambda}{3}  + \frac{g^2}{4}\right) T^2 \\
    \Pi_L &= \frac{g^2}{3} T^2\,.
\end{align}
Finally, we include the standard, properly renormalized Coleman-Weinberg potential~\cite{Coleman:1973jx}. 
At one-loop, we verified that the effects of running the couplings down to the temperature scale are negligible, and that the RG improvement for $\lambda$ leaves us in the perturbative regime.

\bibliographystyle{unsrt}
\bibliography{references}

\end{document}